\documentclass[amssymb,amsfonts,prd]{revtex4}
\pdfoutput=1
\usepackage[dvips]{graphicx}
\usepackage{bm,latexsym,amsmath,amssymb,amsfonts}
\usepackage[usenames,dvipsnames]{color}
\usepackage[colorlinks=true,linkcolor=blue]{hyperref}
\usepackage{soul}
\usepackage[mathscr]{eucal}
\newcommand{\be}[1]{\begin{equation} \label{#1}}
\newcommand{\ee}{\end{equation}}
\newcommand{\bea}{\begin{eqnarray}}
\newcommand{\eea}{\end{eqnarray}}
\newcommand{\ba}{\begin{array}}
\newcommand{\ea}{\end{array}}
\newcommand{\bel}{\begin{align}}
\newcommand{\eel}{\end{align}}
\newcommand{\nn}{\nonumber}
\newcommand{\tcb}{\textcolor{blue}}
\newcommand{\tcr}{\textcolor{red}}

\newcommand{\qq}{K}

\begin{document}

\title{Simple Black Holes with Anisotropic Fluid }
\author{Inyong Cho}
\email{iycho@seoultech.ac.kr}
\affiliation{School of Liberal Arts, Seoul National University of Science and Technology, Seoul 01811, Korea}
\author{Hyeong-Chan Kim}
\email{hckim@ut.ac.kr}
\affiliation{School of Liberal Arts and Sciences, Korea National University of Transportation, Chungju 27469, Korea}
%
\begin{abstract}
We study a spherically symmetric spacetime made of anisotropic fluid of which radial equation of state is given by $p_1 = -\rho$.
This provides analytic solutions and a good opportunity to study the static configuration of black hole plus matter.
For a given equation-of-state parameter $w_2 = p_2/\rho$ for angular directions, we find exact solutions of the Einstein's equation described by two parameters.
We classify the solution into six types based on the behavior of the metric function.
Depending on the parameters, the solution can have event and cosmological horizons.
Out of these, one type corresponds to a generalization of the Reissiner-Nordstr\"{o}m black hole, for which the thermodynamic properties are obtained in simple forms. 
The solutions are stable under radial perturbations.

\end{abstract}
\maketitle

\section{Introduction}

What kind of matter can coexist with a black hole in static equilibrium?
Evidently, an ordinary matter may not be in equilibrium with a classical black hole because of the gravitational attraction near the horizon and the radiation reaction.
Considering quantum mechanical Hawking radiation from the black hole, one may design an equilibrium system~\cite{Tadaki}. 
However, there are well-known examples that matter stays in a stable manner around a black hole, e.g., the charged black hole solution.
For the case of the Reissner-Nordstr\"{o}m solution, the stress-energy tensor of the electro-magnetic field outside the black hole is anisotropic and satisfies $p_1 = -\rho$ in the radial direction and $p_2 = \rho$ in the transverse direction. 
Other exact solution was investigated in Ref.~\cite{Cho:2016kpf}, in which the matter field is assumed to be isotropic with a negative pressure,  $p =  -\rho/3$.
In this case, the energy density vanishes at the black hole event horizon. 
Noting these examples, it is worth while to analyze the solutions of the Einstein's equation having negative radial pressure and an anisotropic configuration to understand equilibrium configurations of the matter and the black hole.

A comprehensive collection of static solutions of Einstein's field equation with spherical symmetry can be found in Stephani {\it et. al.}~\cite{Stephani2003}, Delgaty and Lake~\cite{Delgaty:1998uy}, and Semiz~\cite{Semiz:2008ny}.
Most of them focused on the isotropic fluids because astrophysical observations support isotropy.
The perfect Pascalian-fluid (isotropic-fluid) assumption is supported by solid observational and theoretical grounds.
The Einstein gravity for such isotropic cases has been studied comprehensively, 
and hence the recent attention has been drawn by the isotropic objects in new gravity theories 
such as  the massive gravity. In this theory, the relativistic stars \cite{Katsuragawa:2015lbl}, 
neutron stars \cite{Hendi:2017ibm}, and dyonic black holes \cite{Hendi:2016uni}, for example,
have been investigated very recently.
Other than this tendency, the anisotropic objects in the Einstein gravity have drawn attentions quite a while.
Even though there is no complete consensus that anisotropic pressure plays an important role in compact star, interests on the anisotropic pressure are growing recently~\cite{Isayev:2017hup,Pant:2014dna,Momeni:2016oai,Mak:2002,Mak:2001eb}.
Readers can find other works related with anisotropic matters as following:
in relativistic stellar objects, matter having exotic thermodynamical properties brings anisotropy (see \cite{Herrera1997,Harko:2002db} and references therein).
The local anisotropy in self-gravitating systems were studied~\cite{TOV,Herrera1997}.
The pressure anisotropy affects the physical properties such as the stability and the structure of stellar matter~\cite{Dev2002}.
The self-gravitating charged anisotropic fluid with barotropic equation of state were also considered~\cite{Thirukkanesh2008,Ivanov2002,Varela2010}.
For an Einstein-Maxwell system, anisotropic-charged stellar objects were studied consistent with quark stars in Ref.~\cite{Sunzu:2017coy}.
The Einstein's field equation was solved by assuming some specific mass function in Refs.~\cite{Ratanpal:2016kwu,Matese1980,Mak2003}.
Very recently, the covariant Tolman-Oppenheimer-Volkoff equations for anisotropic fluid was developed in Ref.~\cite{Carloni:2017bck}.

Following these recent trend, in this work, we consider the anisotropic fluid to find exact solutions of the Einstein's field equation.
Adopting the polar coordinates on $S^2$, we write the metric for general spherically symmetric spacetimes as (for derivation of field equations, please see e.g., Ref.~\cite{Chinaglia:2017uqd})
\be{metric}
ds^2 = g_{ab}(x) dx^a dx^b + r^2(x) (d\theta^2 + \sin^2\theta\, d\phi^2),
\ee
where $g_{ab}$ is an arbitrary metric on the two-dimensional Lorentzian manifold $(g_{ab}, M^2)$.
When $g^{ab} (D_a r) (D_b r) \neq 0$, where $D_a$ is the covariant derivative on $M^2$, we can set $r$ as a coordinate on $M^2$ without loss of generality and then the metric may be written as
\be{metric2}
ds^2 = -f(t,r) e^{-\delta(t,r)} dt^2 + f(t,r)^{-1} dr^2 + r^2 (d\theta^2+ \sin^2\theta \,d\phi^2),
\ee
where we do not assume staticity of the spacetime up to now.


The simplifying assumptions mostly used for matter are the vacuum, the electromagnetic field, and the perfect fluid.
For example, the vacuum with the ansatz~\eqref{metric} gives uniquely the Schwarzschild metric~\cite{Birkhoff}, the simplest and best-known black-hole solution.
The stress tensor for an anisotropic fluid compatible with spherical symmetry is  %
\be{st}
T_{\mu\nu} = (\rho+p_2) u_\mu u_\nu + (p_1 - p_2) x_{\mu}x_{\nu} + p_2  g_{\mu\nu},
\ee
where $\rho$ is the energy density measured by a comoving observer with the fluid, and $u^\mu$ and $x^\mu$ are its timelike four-velocity and a spacelike unit vector orthogonal to $u^\mu$ and angular directions, respectively.
The use of this $T_{\mu\nu}$ together with ansatz~\eqref{metric} can describe the interior of static, spherically symmetric, and extremely high-density stars, for example.
In addition to these symmetry, we need to introduce the equation of state, which is the relation between $p_i$ and $\rho$.
In cosmology, one usually assumes the barotropic condition for the equation of state,
\be{eos}
p_i = w_i \rho,
\ee
with $w_i=0$ describing the dust, $w_i=1/3$ the radiation, $w_i< -1/3$ the dark energy, and $w_i< -1$ the phantom energy.
Because the energy-momentum tensor is given by $T^{\nu}_{\mu}= \mbox{diag}(-\rho, p_1,p_2,p_2)$, one of the Einstein equations $G^0_1=0$ gives 
\be{frt}
f(r,t) = f(r).
\ee

Some general requirements for $T_{\mu\nu}$ are proposed, collectively known as {\it energy conditions}.
For example, the weak energy condition states that the energy density should be nonnegative to every observer.
Alternatively, one might impose strong or dominant energy conditions.
For the anisotropic fluid, the energy conditions take the following forms:
the weak energy condition, $\rho \geq 0$, $\rho + p_i \geq 0$,
the strong energy condition,  $\rho+p_i \geq 0$, $\rho + \sum_i p_i \geq 0 $,
the dominant energy condition, $\rho \geq |p_i|$, and
the null energy condition, $\rho + p_i \geq 0$.

In general, $u_\mu$ and $x_\mu$ can be chosen to be arbitrary timelike and spacelike four-vectors.   
Because we are trying to find static solutions, we restrict them to satisfy $u_\mu \propto (\partial_t)_\mu$ and $ x_\mu \propto (\partial_r)_\mu$ in the present work. 
Now, let us consider a matter field across an event horizon described by the fluid form in Eq.~\eqref{st}.
Inside the horizon where $g_{tt}>0$ and $g_{rr}<0$, the coordinate $r$ plays the role of time.
Then, $-p_1$ and $-\rho$ play the roles of the energy density and the pressure along the spatial $t$ direction, respectively.
Upon this switch of the roles, the energy conditions mentioned above do not change and the energy density and the pressure are continuous across the horizon when $w_1 = -1$.
We are interested in this case at the present work.
For other cases, $w_1 \neq -1$, the pressure must be discontinuous at the horizon $r_H$ unless $\rho(r_H) =0$, which implies that solutions satisfying $w_1\neq -1$ and $\rho(r_H) \neq 0$ must be dynamical.
In this work, we require $w_1 = -1$ so that the energy density is continuous across the horizon, which replaces the boundary condition there.
In Ref.~\cite{Delgaty:1998uy}, various tests of acceptability for the isotropic fluid, such as the positivity of energy density and pressure, the regularity at the origin, the subluminal sound speed, etc. were applied to vast of candidates for fluid solutions.

Let us describe the motivation for the anisotropic matter.
The anisotropic fluid can be used to study effectively the static matter fields.
Traversable wormholes are widely studied recently~\cite{Kuhfittig:2018voi,Popov:2017esj} based on various gravity theories. 
The solutions require the existence of exotic materials which violate energy conditions 
and have (effective) negative anisotropic pressures. 
For example, the Morris-Thorne type wormhole satisfies $\rho+ p_1 + 2p_2 =0$~\cite{Morris}.
As will be shown in this work, those materials can also be used to support the matters outside the black-hole horizon. 
For the case of a scalar field, the equation of state varies depending on the kinetic term.
It takes a negative value when the field is non-dynamical,
which can be studied by an anisotropic fluid with negative pressure.
One example is the monopole-black hole in the nonlinear sigma model which will be shown in this work.
For the case of  the static electric field,  the equations of state are given by
$p_1 = -\rho$ and $p_2 = \rho$ and the trace of the stress-tensor vanishes.
This results in the well-known Reissner-Nordstr\"om black hole.
As it was discussed earlier, the condition that the matter stays static at the horizon prescribes $w_1 = -1$ explicitly. 
Therefore, it is worth while to study the solutions of the Einstein's equation 
for anisotropic fluids of  $w_1 = -1$ with various values of $w_2$
as an extension of the Reissner-Nordstr\"om black hole.

In this work, we consider the case of $w_1=-1$ for which there exist exact analytic solutions.
We classify the solutions according to the value of $w_2$.
In Sec. II, we derive the Tolman-Oppenheimer-Volkoff equation for the anisotropic fluid.
In Sec. III, we obtain the exact solution for the case of $w_1 = -1$.
In Sec. IV, we classify the solution into six types and discuss the black-hole types.
In Sec. V, we study the stability of the solutions.
We summarize the results in Sec. VI.

\section{Field equation}
With the metric~\eqref{metric2} with $f(t,r) = f(r)$ as in Eq.~\eqref{frt} and the energy-momentum tensor~\eqref{st}, 
Einstein's equation becomes
\bea
G^0_0 &=& -\frac1{r^2} + \frac{f}{r^2} + \frac{f'}{r} = - 8\pi \rho(r),
    \label{G00}\\
G^1_1 &=& -\frac1{r^2} + \frac{f}{r^2}+ \frac{f'}{r}
	- \frac{f \delta'(t,r)}{r}= 8 \pi p_1(r)
	,
	\label{G11} \\
G^2_2 &=& \frac{f'}{r} + \frac{f''}{2} -
	\frac{f}{2r} \delta'(t,r) -\frac{3f'}{4} \delta '(t,r) +\frac{f}{4} \delta'(t,r)^2 - \frac{f}2 \delta''(t,r)  = 8\pi p_2(r). \label{G22}
\eea
Because we assume $w_1 = -1$, the combination $G^{0}_{~0}- G^1_{~1}=0$ shows $\delta(t,r) = \delta (t)$.
Then, by redefinition of the time coordinate, we can set $\delta =0$ without loss of generality.
Now, the metric has reduced to
\be{metric3}
ds^2 = - f(r) dt^2 + f(r)^{-1} dr^2 + r^2 (d\theta^2+ \sin^2\theta \, d\phi^2).
\ee
This means that there exists a hypersurface-orthogonal Killing vector in the spacetime.
Thus, the spacetime is static in the region where $f>0$, and $\rho=\rho(r)$ and $p_2 = p_2(r)$ hold by consistency.
The first equation~\eqref{G00} can be formally integrated to give
\be{gr}
f(r) = 1-\frac{2m(r)}{r} ,
\ee
where the mass function $m(r)$ is defined by
\be{mr}
 m(r) = 4\pi \int^r r'^2 \rho(r') dr'.
\ee
Here, the integration constant is absorbed into the definition of $m(r)$.
If one requires the analyticity of the spacetime at the center, it requires
$m(r) \simeq  m_3 r^3 + m_5 r^5 + \cdots$ around $r=0$, where $m_3$, $m_5$ are the constants, which restricts the form of $\rho(r)$.
Putting Eq.~\eqref{gr} to Eq.~\eqref{G22}, we obtain the expression of $p_2$ in terms of $\rho$ as
\be{hydro}
p_2 = -\rho - \frac{r\rho'}{2},
\ee
which can also be obtained from the conservation law $\nabla^\mu T_{\mu \nu} =0$.

\section{Analytic solutions}
The purpose of this work is to find analytic solutions of the Einstein's equation.
In this work, we restrict our interests to the exactly solvable case with 
\be{w1}
	w_1=-1.
\ee
When $\rho$ plays the role of an energy density, the energy conditions restrict the matter kinds to physically allowed ones.
Among the conditions, the positivity of energy density appears to be crucial.
In addition to it, all the energy conditions require $ w_2 \geq -1$.
Specifically, the dominant energy condition requires $w_2 \leq 1$ and the strong energy condition requires $w_2 \geq 0 $.
Therefore, when $0\leq w_2 \leq 1$, all the energy conditions will be satisfied.
Once we assume $p_2 = w_2 \rho$, Eq.~\eqref{hydro} is solved to give $m(r)$ for $w_2 \neq 1/2$, the density and the radial pressure 
\be{TOV1}
	m(r) = M + \frac{K}{2 r^{2w_2-1}} , \qquad
\rho (r) = -p_1(r) = \frac{(1-2w_2) K}{8\pi r^{2+ 2w_2}} ,	
\ee
where $M$ and $K$ are constants.
For the energy density to be non-negative, we require
\be{r0}
r_0^{2w_2} \equiv (1-2w_2) K \geq 0,
\ee 
where the positive parameter $r_0$ of  length (mass) scale was introduced for convenience
because the dimension of the parameter $K$ is dependent on the value of $w_2$.
The energy density and the pressure are singular at the origin or at the infinity when $w_2> -1$ and $w_2< -1$, respectively. 
To have a smooth $w_2 \to 1/2$ limit, we introduce a new mass parameter 
\be{rep}
M' \equiv M + \frac{r_0}{2(1-2w_2)}.
\ee
Then, the solutions for $w_2 = 1/2$ can be specified by taking the limit $w_2\to 1/2$ from Eq.~\eqref{TOV1}, which gives  
$$
m(r) = M' + \frac{r_0}{2} \log \frac{r}{r_0} , \qquad \rho(r) = \frac{r_0}{8\pi r^3}. 
$$
All the other physical formulae for $w_2=1/2$ in this work can be obtained with the same manner.
Therefore, we will not discuss the $w_2=1/2$ case separately.  
The metric function in Eq.~\eqref{gr} becomes
\be{fr}
 f(r) = 
  1- \frac{2M}{r} -\frac{K}{r^{2w_2}},
\ee
where $M$ and $K$ can be rewritten by using Eqs.~\eqref{r0} and \eqref{rep}.
Because we are interested in the solutions with matter, we restrict our interests to the case with $r_0 \neq 0$.
For $1/2< w_2 \leq 1$, the spacetime structure must be very similar to that of the Reissner-Nordstr\"om geometry.
For the isotropic fluid with $w_1=w_2=-1$, $M$ and $3K$ represent the mass of the (anti)-de Sitter black hole and the cosmological constant, respectively.
Now, let us analyze the curvature singularities.
The scalar curvature,
$$
R = \frac{ w_2-1 }{r_0^2} \left(\frac{r_0}{r}\right)^{2 (w_2+1)},
$$
is singular at the origin and at the infinity when $-1<w_2  \neq 1$ and $w_2 < -1$, respectively.
Therefore, it is regular everywhere only when $w_2=-1$ and $1$.
The Krestchmann invariant is given by
$$
R_{abcd}R^{abcd} =  \frac{48 M^2}{ r^{6}}+\frac{16 (w_2+1) (2 w_2+1) M r_0^{2w_2}}{ (1-2w_2) r^{2 w_2+5}}+ \frac{4\left(4w_2^4 + 4w_2^3 + 5w_2^2 +1\right)r_0^{4w_2} }{(1-2w_2)^2 r^{4w_2+ 4} }.
$$
Note that the numerator of the last term is positive definite and the apparent divergence for $w_2 = 1/2$ disappears when one introduces the reparameterization in Eq.~\eqref{rep}.
Therefore, the term is singular at the origin or at the infinity when $w_2 > -1$ or $w_2< -1$, respectively.
When $w_2 = -1$, the first term becomes singular at the origin unless $M= 0$.
For $r_0 \neq 0$, the Krestchmann invariant is regular everywhere only when $M=0$ and $w_2 =-1$, of which solution is nothing but the (anti)-de Sitter spacetime.
In summary, the spacetime is singular at infinity when $w_2 < -1$ and singular at the origin when $M \neq 0$.
The singularities can be naked or be covered by (cosmological) event horizons depending on the natures of the spacetime.

Before we discuss the details of each solution,
we would like to discuss the positivity of the energy density
when there exists an event horizon in the solution.
Because $t$ and $r$ coordinates exchange their roles of time and space,
 $-p_1$ plays the role of the energy density in the region of $f(r) < 0$.
Therefore, in the presence of a horizon,
the energy density is positive definite
over the whole space when $w_1 < 0$ and $\rho>0$, which is satisfied with the present solution.

\section{Classification and black hole solutions}
The properties of a solution in Eq.~\eqref{fr} is completely determined by the functional behaviors of $f(r)$. 
We classify the solutions into six types in accordance with the behaviors as following.\\
 (I) Schwarzschild: $f(r)$ changes the signature from negative to positive with $r$. \\
(II) Schwarzschild-de Sitter: $f(r)$ changes the signature from negative to positive and then to negative again.\\
(III) Reissner-Nordstr\"{o}m: $f(r)$ changes the signature from positive to negative and then to positive again.\\
(IV) de Sitter: $f(r)$ changes the signature from positive to negative.  \\
(V) Naked singular: $f(r)$ is positive definite. \\
(VI) Anisotropic cosmological: $f(r)$ is negative definite.
\vspace{5pt}

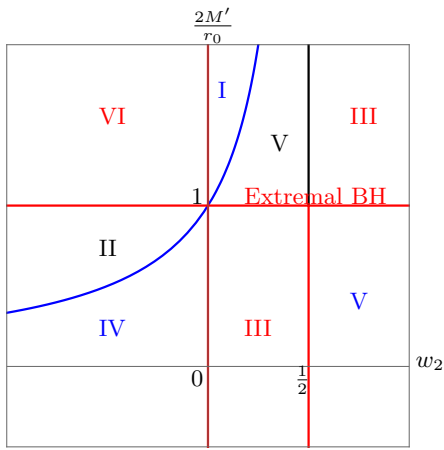
\begin{figure}[bht]
\begin{center}
\begin{tabular}{c}
\includegraphics[width=.3\linewidth,origin=tl]{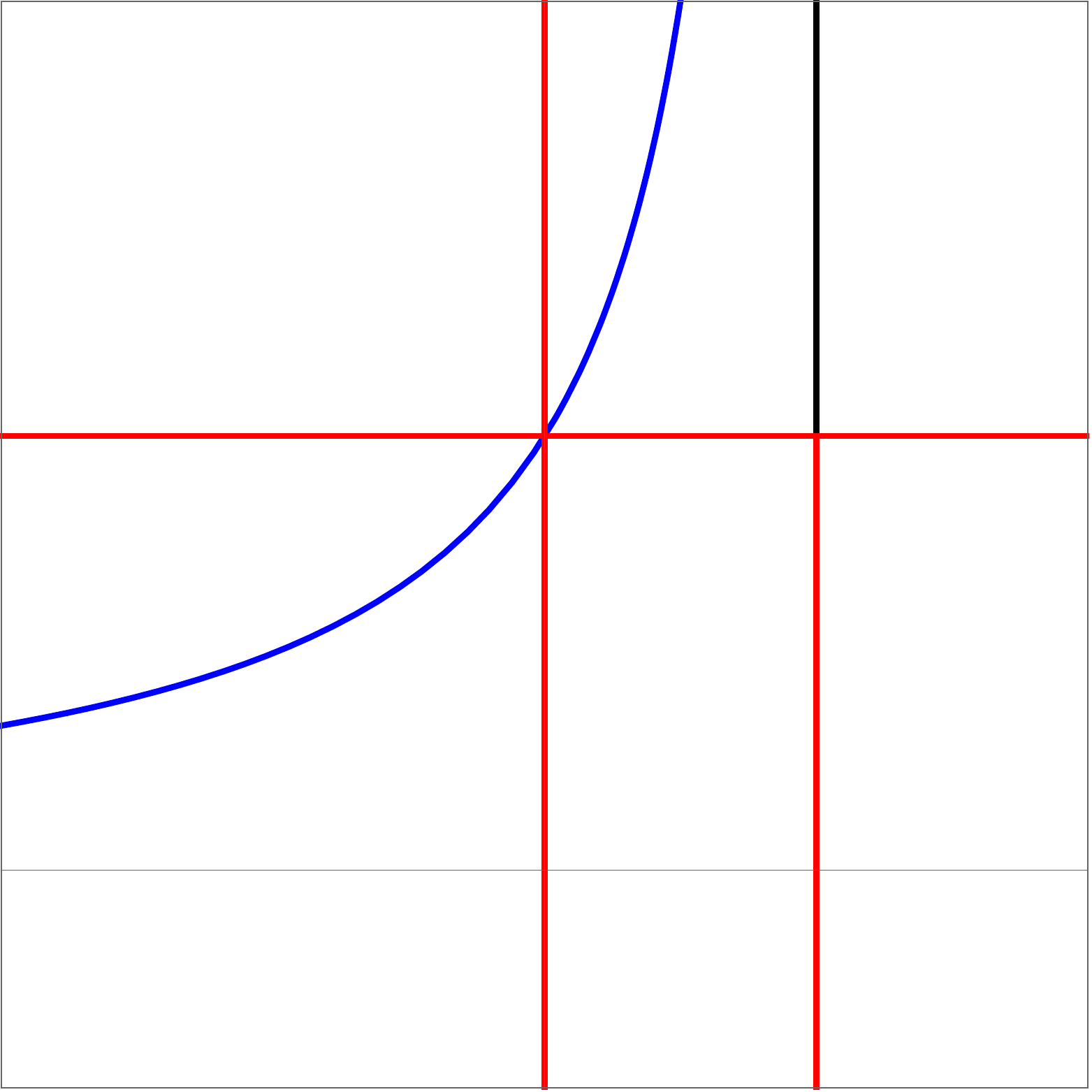}   \\
\end{tabular}
\put( 0,-42) {$w_2$}  \put(-85,85) {$\frac{2M'}{r_0}$}
\put(-46,-49) {$\frac12$} \put(-85,20) {1} \put(-85,-49) {0}
\put(-25,50){\tcr{III}} \put(-25, -20) {\tcb{V}} 
\put(-65,20) {\tcr{Extremal BH}}
\put(-55,40) {V} \put(-65,-30) {\tcr{III}} \put(-75,60) {\tcb{I}}
 \put(-120, 50) {\tcr{VI}} \put(-120,0){II} \put(-120,-30) {\tcb{IV}}
\end{center}
\caption{The classification of the solutions.
The horizontal and the vertical axes denote $w_2$ and $2M'/r_0$, respectively.
The blue curve represents $2M'/r_0 = (1-2w_2)^{-1}$, for which $M=0$.
Each border line in the figure belongs to the part having the same text color.
}
\label{fig:www}
\end{figure}

The full classification of the solutions with respect to the schemes above is given in Fig.~\ref{fig:www}.
The Roman numbers I-VI represent type of the given solution corresponding to the given values of $(w_2, 2M'/r_0)$.

Rather than digress all the details, let us illustrate only the case of $w_2> 1/2$.
Asymptotically, $\displaystyle \lim_{r\to \infty} f(r) \to 1$.
Around the origin,  $f(r)$ is a decreasing function with $f(0) = \infty$.
Its derivative is
$$
f'(r) = \frac{2r_0}{r^2}\left[\frac{M}{r_0} - \frac{w_2}{2w_2-1}\left(\frac{r_0}{r} \right)^{2w_2 -1}\right].
$$
When $M$ is positive, $f'(r)$ vanishes at $r_m = \left[ w_2 r_0/(2w_2-1)M \right]^{1/(2w_2-1)} r_0$ and $f(r)$ has a minimum value at $r_m$,
$$
f(r_m) = 1- \frac{2M}{r_m} + \frac{1}{2w_2-1} \frac{r_0^{2w_2}}{r_m^{2w_2}} 
 = 1- \left[\frac{(2w_2-1)M}{w_2 r_0}\right]^{2w_2/(2w_2-1)}
 = \frac{2w_2-1}{2w_2} \left(1-\frac{2M'}{r_0}\right) \times P
 	,
$$
where $P$ represents a series function of $(2w_2-1)M/w_2 r_0$, which is positive.
The minimum value $f(r_m)$ is positive when $2M'/r_0 <1$.
For this case, the metric function is positive definite for all $r$ and the singularity at the origin is naked.
Now, the metric belongs to type V.
The case of $M<0$ also belongs to this type.
On the other hand, when $f(r_m)<0$ i.e. $2M'/r_0>1$, the spacetime has two horizons at the radius $r_\pm$ satisfying $f(r_\pm)=0$.
Therefore, the spacetime belongs to type III.
When $2M'/r_0 =1$, the two horizons overlap to form an extremal one.

The solutions corresponding to a black hole are type I, II, and III.
The type IV solution is similar to that of the de Sitter spacetime having a cosmological horizon.
On the other hand, the type V solution possesses a naked singularity.
The type VI solution corresponds to a cosmological solution
undergoing anisotropic expansion/contraction which is bounded by a cosmological horizon.
The singularity at $r=0$ is located at the future/past infinity in this case.

\subsection{Type I and II}
The type I solutions exist only when $0\leq w_2 < 1/2$ and $M > 0$.
The geometry takes the form of a modified Schwarzschild spacetime.
The functional form of $f(r)$ at $r\simeq 0$ is governed by the $-2M/r$ term, which implies that $r=0$ is a singularity.
The geometry is asymptotically flat because $f(r)$ approaches a constant value as $r\to \infty$.
There exists an event horizon $r_{\rm BH}$ satisfying $f(r_{\rm BH})=0$.
However, the energy density is not localized sufficiently so that the total mass diverges as $r\to \infty.$

If $w_2=0$, the solution takes the form of a modified Schwarzschild form with a solid angle deficit.
After introducing a new coordinates $y = r/\alpha$ and $\tau = \alpha t$, where $\alpha = \sqrt{1-\qq}$, the metric becomes
$$
ds^2 = - \left(1- \frac{2\tilde{M}}{y} \right) d\tau^2 + \frac{dy^2}{1- 2\tilde{M}/y} + (\alpha y)^2 (d\theta^2 + \sin^2\theta d\phi^2) ,
$$
where $\tilde{M} = M/\alpha^3$.
The horizon is at $r= 2M/\alpha^2$. 
The mass function increases with $r$ as $m(r) = M+ \qq r/2$.
At the horizon, it is $m(r_{\rm BH}) = M/\alpha^2$.
This fluid with $w_2=0$ is equivalent to the self-gravitating global monopole~\cite{Barriola:1989hx},
and the monopole-black hole in the nonlinear sigma model with the hedgehog ansatz~\cite{Gibbons:1990um}.

The type II solution exists only for $w_2 < 0$ with $(1-2w_2)^{-1} < 2M'/r_0< 1$.
The geometry takes the form of the modified Schwarzschild-de-Sitter solution.
The functional form of $f(r)$ at $r\sim 0$ is governed by the $-2M/r$ term, which implies that $r=0$ is a singularity.
Asymptotically, $f(r) \to -\qq r^{2|w_2|}$ goes to negative infinity.
There are two horizons satisfying $f(r) =0$, the one is the black-hole horizon $r_{-}$ and the other is the cosmological horizon $r_+$.
The cosmological horizon resembles that of the de Sitter spacetime.
When $w_2 = -1$, the geometry becomes the Schwarzschild-de Sitter spacetime with
$
f(r) = 1-2M/r - \qq r^2.
$

The mass inside $r$ monotonically increases.
The total mass inside the cosmological horizon becomes 
$$
m(r_+) =\frac{r_+}2\left[\frac{2 M}{r_+} + \frac{1}{1-2w_2} \left(\frac{r_0}{r_+}\right)^{2w_2} \right]
= \frac{r_+}2. 
$$

\subsection{Type III}
The most interesting case is the type III because it contains the Reissner-Nordstr\"{o}m solution as a specific one with the charge $Q = r_0$ when $w_2 =1$.
The solution exists when $2M'/r_0 \leq 1$ and $2M'/r_0 \geq 1$ for $0< w_2 \leq 1/2$ and  $w_2 > 1/2$, respectively.
The geometry of a type III solution takes the form of the Reissner-Nordstr\"{o}m solution in the sense that it has an inner horizon $r_-$ in addition to the outer black-hole horizon at $r_{+}$ given by $f(r_{+}) =0$.
For $w_2  \leq 1/2 $, the mass function grows with $r$. 
Therefore, the solution fails to describe a localized object such as a star.

On the other hand, the matter distribution is localized sufficiently if $w_2 > 1/2$. 
The total mass over the whole spacetime is finite and is given by 
\be{M}
M = M' + \frac{r_0}{2(2w_2-1)} \geq M'.
\ee
Note that $f(r_0) = 1- 2M'/r_0 < 0$, which implies that $r_0$ is located in between the two horizons, $r_- < r_0 < r_+$. 
The surface gravity of the black hole at $r= r_+$ is 
\be{kappa}
\kappa = \frac{f'(r_+) }2 = \frac{1}{r_+} \left[w_2 - \frac{(2w_2-1)M}{r_+} \right]
		 = \frac{1}{2r_+} \left[1- \left(\frac{r_0}{r_+}\right)^{2w_2} \right] \geq 0.
\ee
The entropy and the black hole temperature are given by 
\be{ST}
S = \pi r_+^2, \qquad T_H = \frac{\kappa}{2\pi} = \frac{1}{4\pi r_+} \left[1- \left(\frac{r_0}{r_+}\right)^{2w_2} \right].
\ee
Treating $r_+$ as a function of $M$ and $r_0$ for the variational relation $\delta f(r_+) =0$ in Eq.\eqref{fr}, and using Eqs.~\eqref{M}, \eqref{ST}, one gets the first law of the black hole thermodynamics in the form, 
\be{1st}
T_H \, dS = 2\pi T_H r_+ dr_+ = \delta M -
	\Phi \delta r_0 ,
\ee
where the potential takes the form,
\be{Phi}
\Phi = \frac{w_2 }{2w_2-1} \left(\frac{r_0}{r_+}\right)^{2w_2-1} .
\ee
For the case $w_2 = 1$, this provides the correct value of the electric potential with charge $Q= r_0$.

\section{ stability}
Let us check the stability of the spacetime with anisotropic fluid by introducing linear spherical scalar perturbations.
Let us write the metric in the form,
\be{metric:1st}
ds^2 = - e^\nu dt^2 + e^\lambda dr^2 +e^\mu d\Omega_2^2.
\ee
The general energy-momentum tensor is given by Eq.~\eqref{st} with $p_i = w_i\rho$, where the velocity and the radial four vectors are given by
\be{ux}
u^\mu = [e^{-\nu/2} \sqrt{1+v^2}, e^{-\lambda/2} v,0,0], \qquad
x^\nu = [ e^{-\nu/2} v, e^{-\lambda/2} \sqrt{1+ v^2}, 0,0 ],
\ee
where $v$ corresponds to a normalized radial velocity and $x^a u_a = 0$.
Now, the components of the stress tensor become
\bea \label{T:pert}
T^0_1 &=& (1+w_1) \rho e^{(\lambda-\nu)/2} v\sqrt{1+v^2}, \quad
T^0_0 = - \rho \big[1+(1+w_1) v^2 \big],  \quad
T^1_1 = \rho[w_1+ (1+ w_1) v^2], \quad
T^2_2 = \rho w_2.
\eea

Now, we introduce the perturbations for the metric on a given background solution $\nu_0(r)$, $\lambda_0(r)$, and $\mu_0(r)$ as
\be{pert:geo}
e^\nu = e^{\nu_0(r)}(1 + \delta \nu), \qquad
e^\lambda = e^{\lambda_0(r)}(1 + \delta \lambda),  \qquad
e^\mu = e^{\mu_0(r)}(1 + \delta \mu) , \qquad \rho = \rho_0 + \delta \rho .
\ee
The linearized coordinate transformations of $t$ and $r$ are given by
\be{tr:trans}
t = \tilde t + \delta t(\tilde t, \tilde r), \qquad
r = \tilde r + \delta r(\tilde t, \tilde r).
\ee
Under this transformation, the metric~\eqref{metric:1st} can be written as
\bea
ds^2 &=& - e^{\nu_0(r)}
	(1+ \delta \nu )d t^2 + e^{\lambda_0(r)} (1+ \delta \lambda) d r^2
		+ e^{\mu_0(r)} (1+ \delta \mu) d\Omega^2 \nn \\
&=& 	 - e^{\nu_0(\tilde r)} [1+\delta \nu+ \nu_0' \delta r+ \dot \nu_0 \delta t+ 2\dot{\delta t} ]d\tilde t^2
	+ 2[ e^{\lambda_0} \dot{\delta r} - e^{\nu_0} \delta t']
	d\tilde t d \tilde r
	 \nn \\
&&+ e^{\lambda_0(\tilde r)} [1+ \delta \lambda + \lambda_0' \delta r
	+ \dot{\lambda}_0 \delta t+ 2\delta r'] d\tilde r^2
	+ e^{\mu_0(\tilde r)} \left[1+\delta \mu + \mu'_0 \delta r
		+ \dot \mu_0 \delta t\right] d\Omega^2 , \label{ds2:pert}
\eea
where the overdot and the prime denote the derivatives with respect to $\tilde t$ and $\tilde r$, respectively.
In Eq.~\eqref{ds2:pert}, we can impose $\dot \nu_0 =0$, $\dot \lambda_0 =0$, and $\dot   \mu_0 =0 $.

In this work, we choose the gauge choice for $\delta r$ and $\delta t$ so that
\be{gauge}
\tilde g_{0i}=0 \quad \Rightarrow \quad e^{\lambda_0} \dot \delta r = e^{\nu_0} \delta t', \qquad
 \delta \tilde \mu \equiv \delta \mu + \mu'_0 \delta r=0 .
\ee
Omitting tilde in $\tilde r$ and $\tilde t$, the metric~\eqref{ds2:pert} becomes
\be{ds2:1gf}
ds^2 =  - e^{\nu_0}
	[1+ \delta \nu(t,r) ]d t^2 + e^{\lambda_0} [1+ \delta \lambda(t,r)] d r^2
		+ e^{\mu_0} d\Omega^2 .
\ee
The $\delta G^0_1$ part of the Einstein's equation based on the metric~\eqref{ds2:1gf}, introducing $e^{\nu_0} = f(r) = e^{-\lambda_0}$, determines $v$ to be
\be{G01}
-\frac{\mu_0' \dot {\delta \lambda}}{2 f(r)} = 8\pi (1+ w_1) \frac{\rho_0 v}{f}
\quad \Rightarrow \quad v = - \frac{\mu_0' \, \dot {\delta \lambda}}{16\pi(1+w_1) \rho_0}.
\ee
We may use this equation as a definition of $v$.
The $\delta G^0_0$ part,
\be{derho}
-\frac{f(r)}{4} \left[2\mu_0'\delta\lambda' + \left( 4\mu_0'' + 3\mu_0'^2 + \frac{2\mu_0' f'(r)}{f(r)}\right)\delta \lambda  \right] = -8\pi \delta \rho,
\ee
provides the definition of $\delta \rho$.
The $\delta G^1_1$ part becomes
\be{G112}
\frac{f(r)}{4} \left[2 \mu_0'\delta\nu'  -\left(\mu_0'^2+ \frac{2\mu_0'f'(r)}{f(r)}\right) \delta \lambda  \right] =  8\pi w_1 \delta \rho.
\ee
Combining Eq.~\eqref{derho} and Eq.~\eqref{G112}, we write $\delta \nu'$ by means of $\delta \lambda$ and $\delta \lambda'$:
\be{dnu:dlam}
\delta \nu' = w_1 \delta \lambda'+\Big[2w_1 \frac{\mu_0''}{\mu_0'}+\frac{1+ 3w_1}{2}\mu_0'
		+(1+w_1) \frac{f'(r)}{f(r)}\Big] \,\delta \lambda   .
\ee
Finally,  the $\delta G^2_2$ part becomes
\be{G222}
-\frac{\ddot {\delta \lambda}}{2f(r)} + \frac{f(r)}{4} \left[ 2 \delta \nu'' +
	\big( \mu_0'+ \frac{3f'}{f}\big) \delta\nu'- \Big(\mu_0'+ \frac{f'}{f}\Big)\delta \lambda'
	 - \Big(2\mu_0''+ \mu_0'^2 +\frac{2f''}{f} +\frac{2f'\mu_0'}{f}  \Big) \delta \lambda
	 \right] = 8\pi w_2 \delta \rho.
\ee
Using $\mu_0(r) = 2\log r$ after plugging Eqs.~\eqref{dnu:dlam} and \eqref{G112} into Eq.~\eqref{G222}, we get a differential equation for the perturbation $\delta \lambda$ as
\bea \label{dlam}
\frac{\ddot{\delta \lambda}}{f^2(r)} &=&  w_1 \delta \lambda'' + \left[\frac{2(w_1-w_2)}{r} +
	\frac{1+ 5w_1}2 \frac{f'(r)}{f(r)}\right]\delta \lambda'
	+ V(r) \delta \lambda,
\eea
where
\be{Vr}
V(r) = - \frac{2w_2}{r^2} + \frac{1+ 5w_1- 4w_2}2 \frac{f'(r)}{rf(r)} +
 	\frac{1+w_1}2\frac{f'(r)^2}{f(r)^2} +w_1\frac{f''(r)}{f(r)}.
\ee
Introducing a new coordinate $z$ as $dz = f^{-1} dr$ 
and setting $\delta \lambda =  e^{- i \omega t} g_1(r)$, 
the perturbation equation becomes
\be{pert}
- \frac{d^2 g_1}{dz^2} + \left[\frac{2(w_1-w_2)}{w_1 r}f(r) +
	\frac{1+ 3w_1}{2w_1} f'(r)\right] \frac{dg_1}{dz} + V(z) g_1
		= \frac{\omega^2}{w_1} g_1,
\ee
where $V(z) = V(r(z))$.
For large $\omega^2$, the equation takes the form, $d^2g_1/dz^2 \approx (-w_1^{-1}) \omega^2 g_1$.
Therefore, the system has possibility for being stable only when $w_1 \geq 0$.

An interesting exception is the present case with $w_1 = -1$.
The off-diagonal component of the stress tensor $T_1^0$ in Eq.~\eqref{T:pert} vanishes, 
and $\delta \lambda$ is independent of time from Eq.~\eqref{G01},
\be{deltalambda}
\dot{\delta\lambda} = \ddot{\delta\lambda} =0.
\ee
Then, Eq.~\eqref{dlam} becomes, introducing $\Delta = f \delta \lambda$, 
\be{30}
\Delta '' + \frac{2(1+w_2)}{r} \Delta' + \frac{2w_2}{r^2} \Delta - (1+w_1) \left[
	\Delta'' + \left(\frac{f'}{2f}+ \frac2r\right) \Delta' + \frac{f'}{2rf} \Delta\right] =0 .
\ee
For $w_1 =-1$, ignoring the term in the square-bracket, the equation allows an exact solution,
\be{lambda11}
\delta \lambda =\frac1{f(r)} \left(
   	 \frac{2\delta M}{r} + \frac{\delta K}{r^{2w_2}}  \right) ,
\ee
where $\delta K$ represents the variation of $K \equiv r_0^{2w_2}/(1-2w_2)$. 
Then the metric component $g_{11}$ in Eq.~\eqref{ds2:1gf} becomes
\begin{equation}
g_{11}^{-1} = e^{-\lambda_0} [1+ \delta \lambda(t,r)]^{-1}
\approx f(r) \left( 1-\delta\lambda \right)
=
   	 1- \frac{2(M+\delta M)}{r} - \frac{K+\delta K}{r^{2w_2}}  . 
\end{equation}
This corresponds simply to a redefinition of the parameters $M$ and $\qq$. 
The perturbations do not cause an instability for $w_1=-1$.
Therefore, the solution is stable under radial perturbations.
There are various types of solutions in the present model depending on the values of $w_2$.
It is interesting that all of them are stable under the perturbations.

\section{Summary}
We have studied static geometries with an anisotropic fluid based on Einstein's theory of gravity.
Because of the spherical symmetry, the angular pressures should be isotropic.
However, it can be different from the radial pressure without breaking the spherical symmetry.
When $p_1 = -\rho$, the Einstein equation takes simple form,
$p_2 = -\rho - r\rho'/2,$
of which solution allows various classes with $p_2= w_2 \rho$.

Formally, by using the areal radial coordinate, the solution takes similar metric with that of the Schwarzschild-de Sitter-like spacetime,
$$
g_{rr}^{-1} = g_{tt} = 1- \frac{2M}{r} -\frac{\qq}{r^{2w_2}} 
.
$$
After reparameterizing the parameters as $M \equiv M' - r_0/2(1-2w_2),$ and $ \qq \equiv r_0^{2w_2}/(1-2w_2) $, we have shown that the spacetime geometry is determined by one parameter $2M'/r_0$ for the given equation-of-state parameter $w_2$.
For the isotropic case, they represents the mass and the cosmological constant.
When $M \neq 0$, the solution has a curvature singularity at $r=0$.
It has another singularity at infinity when $w_2 < -1$ and $r_0 \neq 0$.
On the basis of the behavior of the metric function, we classified the solutions into six types, including black holes (types I, II, and III), de Sitter-like (type IV) and anisotropic cosmological type VI solutions.
Of all the solutions, the type III black hole solution with $w_2 > 1/2$ describes a physically relevant localized object.
The thermodynamics of the black hole was shown to have the usual form with a potential $\Phi = w_2/(2w_2-1) \times (r_0/r_+)^{2w_2-1}$ and the corresponding charge $r_0$.

We have also developed the stability analysis for anisotropic fluids.
For the $w_1 = -1$ case, we found that the developed instability is a gauge artifact.
Therefore, the solution is stable under radial perturbations.
There are various types of solutions in the present model.
It is interesting that all of them are stable under the perturbations.
We found the fluid configurations coexisting with a black hole in static equilibrium.
Such stable static black-hole solutions are rare, so
further investigation of these solutions are to be performed in the future.

\section*{Acknowledgment}
This was supported by Korea National University of Transportation in 2017 and 
by the grants from the National Research Foundation 
funded by the Korean government
No. NRF-2017R1A2B4010738 (I. C.).

\end{document}